\documentstyle[epsfig,dina4]{article}
\newcommand{\p}{\mbox{e$^+$}}
\newcommand{\n}{\mbox{e$^-$}}
\newcommand{\pn}{\mbox{e$^+$e$^-$}}
\newcommand{\UTa}{\mbox{$^{238}$U + $^{181}$Ta\ }}
\newcommand{\UPba}{\mbox{$^{238}$U + $^{208}$Pb\ }}
\newcommand{\UPb}{\mbox{$^{238}$U + $^{206}$Pb\ }}
\newcommand{\Pba}{\mbox{$^{206}$Pb\ }}
\newcommand{\Pbb}{\mbox{$^{207}$Pb\ }}

\begin{document}
\begin{titlepage}
\hspace*{12.5 cm}{\bf UCY--PHY--96/17}
\vspace*{0.9cm}
\begin{center}
{\huge \bf \mbox{\pn{} pairs from \UPb collisions}} 
\end{center}
 
\vspace*{1.1cm}
\begin{center}
\vspace{0.3cm}
{\Large
 S.~Heinz$^a$, E.~Berdermann$^b$, F.~Heine$^a$,   
 O.~Joeres$^a$, P.~Kienle$^a$, I.~Koenig$^b$, W.~Koenig$^b$,
 C.~Kozhuharov$^b$, U.~Leinberger$^b$, 
 M.~Rhein$^b$, A.~Schr\"oter$^b$, H.~Tsertos$^{c,}$\footnote{
  Corresponding author, e-mail ''tsertos@alpha2.ns.ucy.ac.cy'' \\
  Dept. Nat. Science, Univ. of Cyprus, PO 537, 1678 Nicosia, Cyprus}
}

\vspace{0.7cm}
{\large \bf (The ORANGE Collaboration at GSI)}

\large
\vspace{0.3cm}
$^a$ Technical University of Munich,  D--85748 Garching, Germany\\
$^b$ Gesellschaft f\"ur Schwerionenforschung (GSI),
                                              D--64291 Darmstadt, Germany\\
$^c$ University of Cyprus, CY--1678 Nicosia, Cyprus\\
\end{center}

\large
\vspace*{1.0cm}
\begin{center}
\section*{Abstract}
 \end{center}
We present the results from \pn--coincidence measurements in \UPb 
collisions at a beam energy of 5.93 MeV/u, using an improved experimental 
setup at the double-Orange spectrometer of GSI.
The capability of our device to detect 
Internal-Pair-Conversion (IPC) \pn{} pairs from discrete nuclear transitions
of a moving emitter is demonstrated 
by investigating the known 1.844 MeV (E1) transition in \Pba  and 
the 1.770 MeV (M1) transition in \Pbb, excited via Coulomb excitation 
and a neutron transfer reaction, respectively.
The Doppler-shift corrected \pn--sum-energy spectra show weak lines at the 
expected energies with cross sections being consistent with the measured
excitation cross sections of the corresponding $\gamma$ lines 
and the theoretically predicted IPC coefficients. 
No other \pn--sum-energy lines were found in the measured spectra.
\noindent
The observed transfer cross sections show a strong dependence on the 
distance of closest approach (R$_{min}$), thus signaling also a strong 
dependence on the bombarding energy close to the Coulomb barrier.
\end{titlepage}

\section{Introduction}

\large
Previous results of the EPOS and ORANGE collaborations
at the UNILAC accelerator of GSI have revealed 
unexpected lines in the \pn--sum-energy spectra obtained 
in heavy-ion collisions at bombarding energies near the 
Coulomb barrier~\cite{Sal90,Ikoe93}.
No viable explanation could be found for these experimental results. 
At the beginning it was tempting to interpret these lines as being
due to the \pn{} decay of a previously unknown neutral particle with a mass
around 1.8 MeV/c$^2$, a conjecture which was definitely 
ruled out by subsequent conclusive Bhabha-scattering 
experiments~\cite{Tse91}.   
After this time it was become clear that Internal Pair Conversion (IPC)
from excited nuclear transitions, which can in principle lead to narrow
lines in the \pn--sum-energy spectra, was less well understood than 
previously assumed, such that some of reported weak lines
might be due to this process.

\vspace*{2mm}
\hspace*{3mm}
The motivation of the present work was indeed a search for discrete 
IPC transitions, excited in heavy-ion collisions at energies close to the
Coulomb barrier, as possible candidates for previously observed weak
\pn-sum-energy lines with cross sections of the order of a few $\mu$b for the
\UTa  system and some tenth of $\mu$b for $^{238}$U + $^{208}$Pb 
collisions~\cite{Ikoe93}. 
Particularly for the last system, two
\pn--sum-energy lines at (575 $\pm$ 6) keV and at (787 $\pm$ 8) keV 
have been observed at a beam energy of 5.9 MeV/u~\cite{Ikoe93}. 
The 787 keV line
appeared in both quasielastic central (R$_{min} <$ 20 fm) and 
rather peripheral (R$_{min} =$ 20--26 fm) collisions, whereas
the 575 keV line occurred only in coincidence with central collisions.
Both lines were seen without a Doppler-shift correction and only
in the opening-angle region $\theta_{\pn}$= 
$155^{\circ} - 177^{\circ}$, for which Doppler broadening is expected to be
small for similar energies of both leptons. 

\vspace*{2mm}
\hspace*{3mm}
In order to test the response and sensitivity
of our experimental setup to IPC pairs from a moving emitter we investigated
a known transition in $^{206}$Pb populated via Coulomb excitation, using a 
$^{238}$U beam and a $^{206}$Pb target. 
On the other hand, since the number of combined nuclear charge (Z$_u=$174)
is the same for the collision systems \UPb and \UPba, the \pn{} lines seen
previously in the last system~\cite{Ikoe93} should also appear
in the present experiment, if their origin is closely connected with 
the strong electromagnetic fields available in the collision.

\vspace*{2mm}
\hspace*{3mm}
Here we report the first results from these improved investigations, being
performed at the UNILAC accelerator of GSI.
A dedicated Doppler-shift technique is exploited 
at the double-Orange setup~\cite{Lein96a}, which allows to reveal 
narrow lines in the measured
\pn--sum-energy and $\gamma$-ray spectra~\cite{Shei96}.

\vspace*{2mm}
\hspace*{3mm}
It should be mentioned here that, from their experimental results 
reported most recently, the APEX collaboration~\cite{APEX95} at ANL and 
the EPOS II~\cite{Gan96} and ORANGE~\cite{Lein96b} collaborations at 
GSI have 
all failed to find evidence for the previously reported \pn--sum-energy lines.

\vspace{0.5 cm}
\section{Experimental setup}

As shown in Fig. 1, leptons
emitted from a target placed between two toroidal magnetic field
spectrometers with their axis parallel to the beam direction were momentum
analyzed by the toroidal magnetic field and detected with two arrays 
of high-resolution Si detectors.
The toroidal $(\frac{1}{r}$)-field is generated by 60 iron-free coils. 
Electrons with emission angles $\theta_{\n} = 38^{\circ} 
- 70^{\circ}$ and positrons with $\theta_{\p} = 110^{\circ} - 145^{\circ}$ 
relative to the beam axis are accepted by the spectrometers. 
The lepton detectors consist each of 72 Si PIN diodes (chips)
of trapezoidal shape (base: 24 mm, top: 16 mm, height: 16 mm, thickness: 1 mm) 
arranged in a Pagoda-like form~\cite{Lein96a}. 
Each chip is subdivided into three segments.
A Pagoda roof consists of six PIN diodes, i.e. 18 segments. One detector array
is composed of 12 such roofs. 
Thus, we get a position sensitive detector with
216 segments read out in a matrix mode. At a given field setting only 
particles with a certain sign of charge are
focused onto the corresponding detector arrays. 
Thus, a very clean separation
of electrons and positrons is achieved. For further lepton identification the 
lepton energy and momentum is determined simultaneously. The 
deposited energy is measured by the PIN diodes, and the lepton 
momentum is calculated from the deflection and the
field setting. Only events for which the energy-momentum relation is fulfilled
are accepted, with the result that the remaining e$^+$ misidentification is
small and can be determined reliably (see also Refs.~\cite{Lein96a,Lein96b}). 

\vspace*{2mm}
\hspace*{3mm}
The spectrometer accepts lepton pairs with opening 
angles, $\theta_{e^+e^-}$, from  40$^{\circ}$ to $180^{\circ}$ in the 
laboratory system. 
This range 
can be subdivided into 10 bins of width $\approx \pm 10^{\circ}$. 
For reconstruction of the reaction kinematics both scattered heavy ions are 
detected by 19 Parallel Plate Avalanche Counters (PPACs) which accept ions
scattered under polar angles of $\theta_{ion} = 12.5^{\circ} - 35^{\circ}$ 
and $\theta_{ion} = 40^{\circ} - 70^{\circ}$ with a resolution of 
$1.0^{\circ}$ and $0.5^{\circ}$, respectively. 
The azimuthal-angle resolution of all (ion and lepton) detectors is 
$\Delta\phi=20^{\circ}$. \\
For the detection of $\gamma$ rays we used a high resolution
70\%-Ge(i) detector placed
at $\theta_{\gamma} = 86^{\circ}$ relative to the beam axis at a 
distance of 40 cm from the target.
An ionization chamber installed at $\theta = 40^{\circ}$ measures 
the energy of scattered particles and thus controls the effective target
thickness. It is also used for current normalization.  

\vspace*{2mm}
\hspace*{3mm}
Extensive measurements carried out with
radioactive $^{90}$Sr and $^{207}$Bi sources proved that our setup is
capable of detecting IPC pairs and of determining their opening-angle
distribution from an emitter at rest~\cite{Lein96a}.
The measured FWHM of these
sum-energy lines is $\sim$16 keV, consistent with the sum-energy resolution 
of the lepton detectors.

\vspace{0.5 cm}
\section{Experiment and results}

Data were taken for the collision system $^{238}$U + $^{206}$Pb using
$^{238}$U beams and 800 $\mu$g/cm$^2$ thick $^{206}$Pb 
targets mounted on a rotating target wheel. 
The projectile energy (5.93 MeV/u) is slightly below the Coulomb 
barrier (6.06 MeV/u). The $3^-$-level at 2.65 MeV
in $^{206}$Pb is populated via Coulomb excitation and deexcites
for the most part into the lower lying $2^+$-level with a 
$\gamma$-transition energy of \mbox{1844 keV}~\cite{Led78}.
Following an event-by-event Doppler-shift correction to the Pb-like 
recoiling ion, a pronounced $\gamma$ line with a total excitation 
cross section of $\sigma_{\gamma}=$\mbox{(55 $\pm$ 5) mb} appears at 
(1844 $\pm$ 1) keV in the measured energy spectrum (Fig.2a). 

\vspace*{2mm}
\hspace*{3mm}
The excitation probability, P$_{\gamma}$(R$_{min}$), was determined as
a function of distance of closest approach, R$_{min}$, by normalizing the
$\gamma$ yield in certain R$_{min}$ intervals with the corresponding number
of elastically scattered ions. 
At large R$_{min}$ values, P$_{\gamma}$ shows an
exponential decrease, whereas at small R$_{min}$ values P$_{\gamma}$ is cut off
abruptly at R$_{min} \sim$ 17 fm, where the nuclei come into contact.
Such a behaviour is typical for Coulomb excitation (Fig. 2b).
The results are in accordance with measurements of the EPOS collaboration who 
investigated the same collision system at a bombarding energy of 5.82 MeV/u and
$\sim$400 $\mu$g/cm$^2$ thick targets. They report recently a measured 
cross section of
$\sigma_{\gamma}=$ (44 $\pm$ 7) mb~\cite{Bau96}.

\vspace*{2mm}
\hspace*{3mm}
From the 1844 keV transition we expect IPC pairs with a sum energy of 
\mbox{822 keV.} Multiplying the measured $\gamma$ cross section 
with the theoretically predicted IPC coefficient of 
$\alpha_{IPC} = 4.0 \times 10^{-4}$, for an electromagnetic
transition with multipolarity E1, Z=82 and energy 1850 keV~\cite{Schl79}, we
expect a total cross section for IPC of $\sigma_{IPC}=$(22 $\pm$ 2) $\mu$b. 
In order to optimize
the peak-to-continuum ratio in the e$^+$e$^-$-sum-energy spectra, we eliminated
very positive energy differences $\Delta$E = E$_{e^+} -$ E$_{e^-} >$ 175 keV 
and accepted only rather peripheral collisions with R$_{min} \geq$ 23 fm. 
Thus, we could reduce the contribution of the continuum pairs produced by 
the large time changing Coulomb field.
The event-by-event Doppler-shift corrected \pn--sum-energy spectrum 
obtained under these conditions is shown in Fig. 2c. 
The spectrum is integrated over the whole range of lepton opening angles
covered experimentally.
As shown, the
continuous part of the measured spectrum is well reproduced by a reference
distribution (smooth solid curve), being gained by an event-mixing procedure. 

\vspace*{2mm}
\hspace*{3mm}
From the IPC cross section quoted above, a total of about 35 IPC pairs is 
then expected in the spectrum shown in Fig. 2c, using an IPC detection 
efficiency 
of $\epsilon_{IPC} =$ (1.6 $\pm$ 0.2) $\times 10^{-3}$.
They should result in a line at an energy of $\sim$820 keV with a FWHM 
of $\sim$ 20 keV, superimposed on the \pn{} continuum.
The IPC detection efficiency has been obtained by a Monte Carlo 
simulation, which assumes
isotropically emitted e$^+$e$^-$ pairs and theoretically-calculated 
lepton energy distributions for Z $=$ 82 and E1 multipolarity~\cite{Hof}.
The width of the Doppler-shift corrected line depends
on the emitter velocity and on the angular resolution of the detectors. 
For the case of a rather slow emitter (i.e. $\approx$0.05c), we expect 
line widths near the limit given by the detector sum-energy resolution,
i.e. FWHM $\sim$ 16 keV~\cite{Lein96a}.
As can be seen in Fig. 2c, at an energy of (815 $\pm$ 10) keV an excess 
of (34 $\pm$ 12) counts is observed, in accordance with our estimate 
of 35 counts.

\vspace*{2mm}
\hspace*{3mm}
We found an additional line in the $\gamma$ spectra, when corrected on 
the Pb-like scattered ion, at an energy of (1770 $\pm$ 1) keV.
This line appears only in central collisions with R$_{min} <$ 20 fm 
(Fig.3.a). 
In the R$_{min}$ parameter range selected
(17 fm $<$ R$_{min}$ $<$ 20 fm), the $\gamma$ lines at 1770 keV and at 
1844 keV have comparable intensities. 
The excitation probability for the 1770 keV line
as a function of R$_{min}$, is shown in Fig. 3.b. 
It peaks within a narrow R$_{min}$ interval, typical for transfer reactions, 
for which the transfer probability is expected to become large when the 
nuclei come into contact. 
This also means 
that the excitation function should exhibit a rather narrow structure at
energies close to the Coulomb barrier with a width of about 0.3 MeV/u. 
Similar narrow structures in the R$_{min}$ dependence were observed for a
two-neutron transfer reaction, leading to the known 3.71 MeV $5^-$ level 
in $^{208}$Pb~\cite{Led78}.

\vspace*{2mm}
\hspace*{3mm}
The line at 1770 keV is assigned to a known M1 transition in $^{207}$Pb
from the $\frac{7^-}{2}$ (2.34 MeV) to $\frac{5^-}{2}$ (0.57 MeV) 
state~\cite{Led78}, populated by neutron transfer 
from $^{238}$U to $^{206}$Pb, with a total
cross section of $\sigma_{\gamma}=$(1.1 $\pm$ 0.3) mb. 
The corresponding IPC production cross
section is expected to be 
$\sigma_{IPC}=$(0.3 $\pm$ 0.1) $\mu$b with 
$\beta = 2.8 \times 10^{-4}$~\cite{Schl79}, which should lead to a weak 
\pn-sum-energy line at $\sim$750 keV  
with a FWHM of $\sim$ 40 keV, and an intensity of about 10 counts. 
This is also weakly indicated in the corresponding Doppler-shift corrected
\pn--sum-energy spectrum, shown in. Fig. 3c, although one cannot distinguish  
its appearance from the expected statistical fluctuations, which are of 
the same order. 
The situation is similar for the line at $\sim$820 keV, which is
expected to have comparable intensity.

\vspace*{2mm}
\hspace*{3mm}
Note that the 1770 keV $\gamma$ line was not observed by the EPOS 
collaboration, who investigated the same collision system 
at a beam energy of 5.82 MeV/u and
a $\sim$ 400 $\mu$g/cm$^2$ thick $^{206}$Pb target~\cite{Bau96}.
This may also be taken as an indication for a strong
beam energy dependence of transfer reactions at the Coulomb barrier of
heavy collision systems.

\vspace{0.5 cm}
\section{Summary and Conclusions}

The results discussed in the previous section show that IPC processes after
Coulomb excitation appear with typical cross sections of some $\mu$b in the 
e$^+$e$^-$--sum-energy spectra. Corresponding $\gamma$ lines resulting from 
Coulomb excitation have cross sections of some 10 mb.
This is expected for IPC coefficients of the order of $10^{-4}$. 
The observed cross sections of the
$\gamma$ lines, following nucleon transfer, show a peaked R$_{min}$ dependence
and are in the order of 1 mb. They suggest corresponding e$^+$e$^-$--line cross 
sections of some 0.1 $\mu$b, in agreement with the measurement.

\vspace*{2mm}
\hspace*{3mm}
Previously reported e$^+$e$^-$--sum-energy lines, showing partially the
characteristics of IPC transitions in the Ta-like 
nucleus~\cite{Lein96a}, reveal cross
sections of a few $\mu$b. From this limit, $\gamma$ 
transitions of at least 10 mb cross section are suggested. 
However, none of the corresponding 
Doppler-shift corrected $\gamma$-ray spectra exhibited 
lines with cross sections $\sigma_{\gamma} >$ 1 mb, thus
excluding electromagnetic transitions with multipolarity $\ell >$ 0 in 
the Ta-like nucleus as the source of the \pn{} lines. 
However, transitions without a $\gamma$
decay branch, namely $0^+ \rightarrow 0^+$ transitions, cannot be excluded
by the $\gamma$ measurements.

\vspace*{2mm}
\hspace*{3mm}
Recently, a search for monopole transitions in \UTa collisions 
at a beam energy of 6.0 MeV/u
was conducted at GSI using conversion electron spectroscopy~\cite{Die96}. 
From these investigations, a limit of 0.3 mb for the production cross section 
of a K-conversion line in $^{181}$Ta at an energy around 1.7 MeV is derived.
This limit can be transformed into an upper limit 
of $\sigma_{IPC} \leq 10$ $\mu b$
for the production of corresponding \pn{} pairs, when an     
IPC-conversion-electron coefficient of  $\eta = 0.03$~\cite{Soff81} for 
$^{181}$Ta at an energy of 1.8 MeV is taken into account.
In the case of our previously reported lines~\cite{Lein96a}, this limit is not
quite sensitive enough to exclude an E0 transition definitely.

\vspace*{2mm}
\hspace*{3mm}
In context with the 575 keV and 787 keV \pn--sum-energy lines, observed 
previously in $^{238}$U + $^{208}$Pb collision at 5.9 MeV/u~\cite {Ikoe93}, one
would expect 
$\gamma$ lines at 1.595 MeV and 1.807 MeV energy with cross sections of
about 1 mb, assuming the $^{208}$Pb-like nuclei as emitters.
Unfortunately, the corresponding $\gamma$-ray spectra were taken at that 
time with
a low-resolution NaI detector~\cite{Ikoe93},  which was not sensitive 
enough for detection of such weak $\gamma$ lines. 
Consequently, no exclusion of electromagnetic transitions of
any multipolarity is possible at this stage for this collision system.

\vspace*{2mm}
\hspace*{3mm}
In summary we conclude that IPC \pn{} pairs from discrete nuclear transitions
of a moving emitter can be observed with production
cross sections down to the tenth $\mu$b level in heavy collision 
systems, in which the 
continuous distributions due to uncorrelated \pn{} emission and 
nuclear background are dominating.
Apart from the observed weak IPC lines, no other lines could be found
in the measured \pn--sum-energy spectra of this collision system.
The present sensitivity achieved is at the level of
statistical fluctuations or weak \pn--sum-energy 
lines observed in previous experiments. 

\vspace*{1.5cm}
\large
{\bf Acknowledgement:}
{\em We would like to thank all the people of the UNILAC accelerator
 operating crew for their efforts in delivering stable $^{238}$U
 beams with high intensities.}

\newpage

\centerline{\Large \bf Figure Captions}                              

\vspace{1.5cm}
\noindent
{\bf Fig. 1.} Schematic view of the ORANGE spectrometers. The setup consists of
the following components: two Si-(PIN)-diode arrays (PAGODAs) for lepton
detection, 19 PPACs to count the scattered heavy ions, an intrinsic Ge 
detector for $\gamma$-ray detection and a rotating target wheel.

\vspace{1.0cm}
\noindent
{\bf Fig. 2.} 
{\bf a)} Spectrum of $\gamma$ rays from \UPb  collisions after 
Doppler-shift correction to the Pb-like ions, scattered in the R$_{min}$ 
range from 17 to 32 fm. The line at (1844 $\pm$ 1) keV belongs
to the E1 transition 3$^-$ (2.65 MeV) 
$\rightarrow$ 2$^+$ (0.80 MeV) in $^{206}$Pb.\\ 
{\bf b)} The excitation probability of the 1844 keV 
$\gamma$ transition as a function of R$_{min}$.\\
{\bf c)} The Doppler-shift corrected \pn--sum-energy spectrum obtained 
under the assumption that Pb-like nuclei are the emitters 
for R$_{min}=$ 23 --32 fm
and \pn{} energy differences from $-$200 to 175 keV. The contribution of
random coincidences ($\sim$ 15\%) is subtracted from the data.
The smooth solid line is a reference continuous distribution gained by event
mixing.

\vspace{1.0cm}
\noindent
{\bf Fig. 3.} {\bf a)} Doppler-shift corrected $\gamma$-ray spectrum for 
rather central collisions (R$_{min} \leq$ 20 fm). 
The line at (1770 $\pm$ 1) keV corresponds to the M1 transition 
$\frac{7^-}{2}$ (2.34 MeV) $\rightarrow$
$\frac{5^-}{2}$ (0.57 MeV) in $^{207}$Pb, produced 
 by 1n-transfer reaction.\\
{\bf b)} Excitation probability as a function of R$_{min}$ of the 1770 keV
$\gamma$ transition. \\
{\bf c)} The same as in Fig. 2c, but for R$_{min}$ values between 
   17 and 20 fm.
%

\newpage
\pagestyle{empty}

\vspace*{4.5cm}
\begin{center}
 \epsfig{file=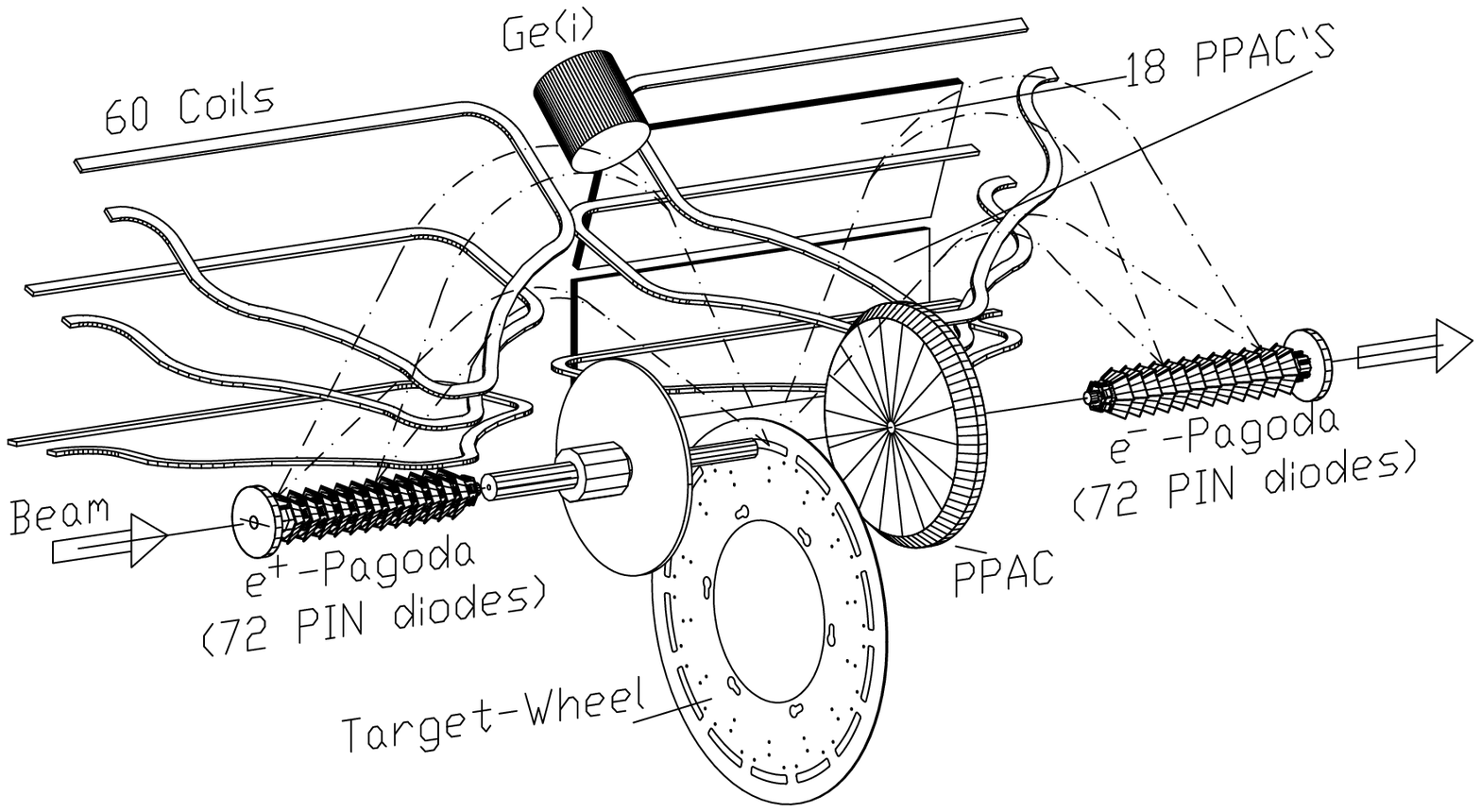,width=1.1\linewidth}
\end{center}

\vspace*{5.0cm}
{\Large \bf Figure 1}

\newpage 
\pagestyle{empty}

\begin{center}
 \epsfig{file=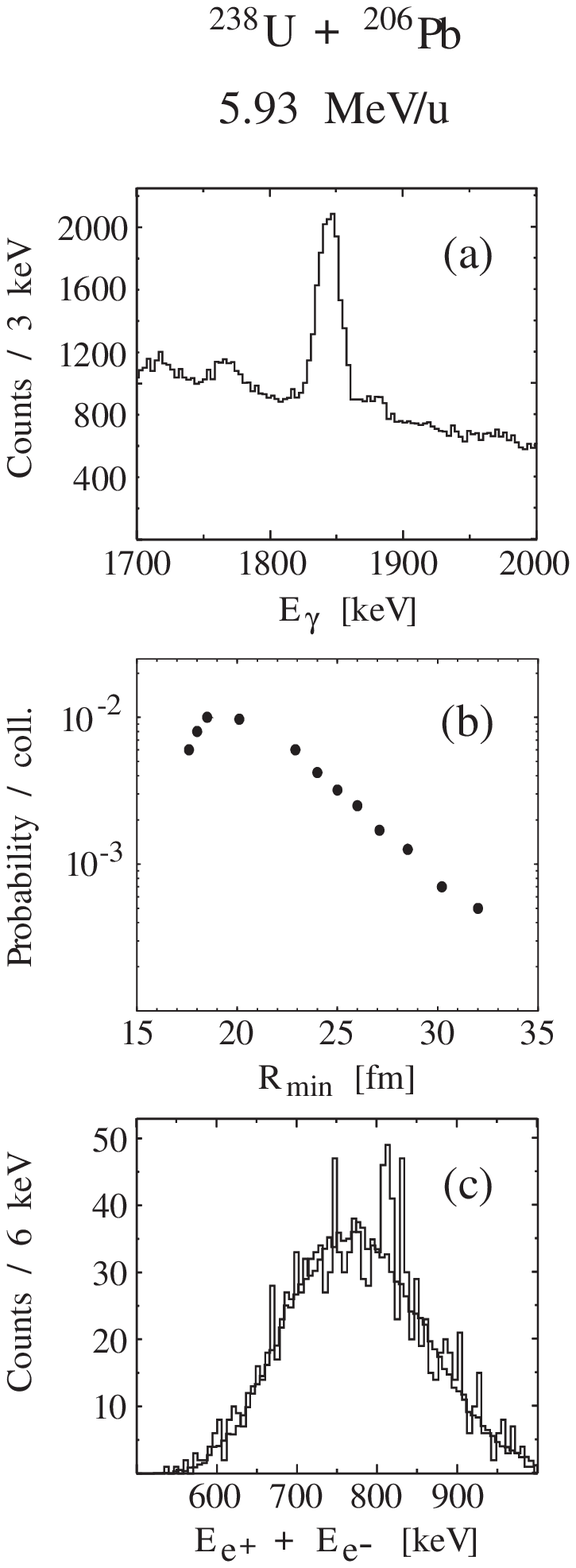,height=21 cm}
\end{center}

\vspace*{0.4cm}
{\Large \bf Figure 2}

\newpage
\pagestyle{empty}

\begin{center}
 \epsfig{file=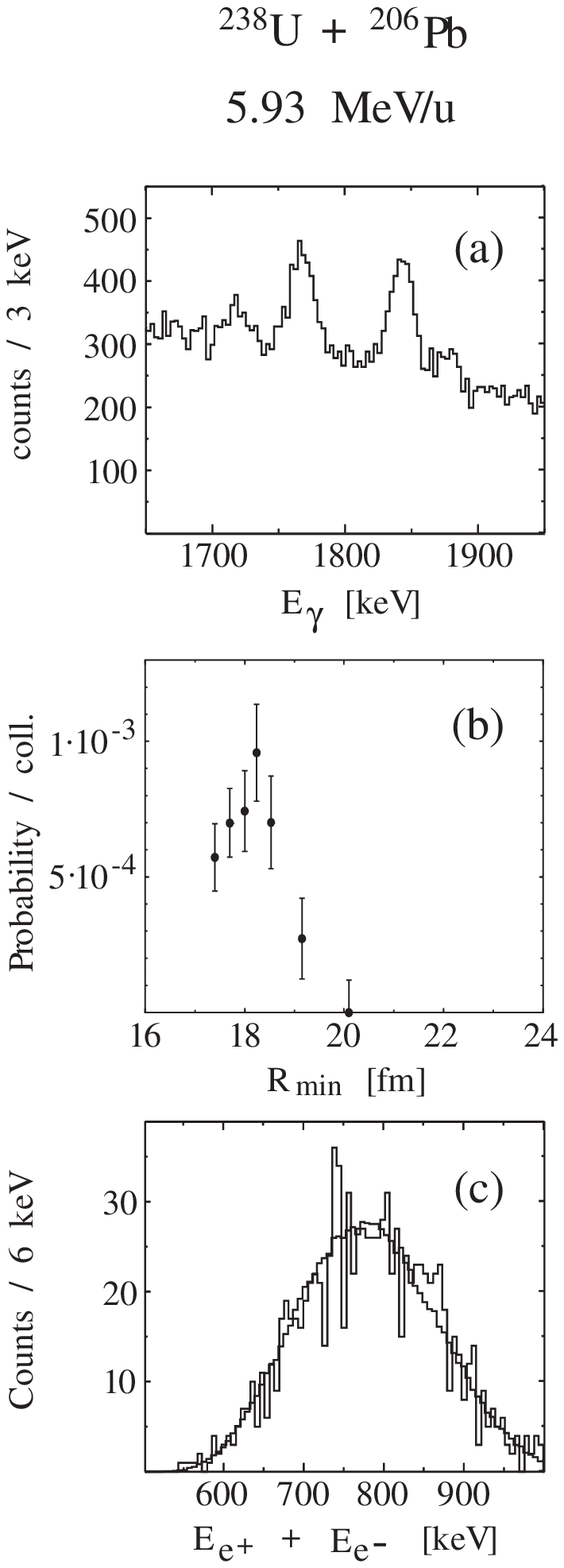,height=21 cm}
\end{center}

\vspace*{0.4cm}
{\Large \bf Figure 3}


\newpage
\vspace*{4mm}
\begin{thebibliography}{99}

\bibitem{Sal90}
 EPOS collaboration, P. Salabura {\em et al.}, Phys.  Lett.  
{\bf B 245}, 153 (1990); see also
T.~Cowan {\em et al.}, Phys. Rev. Lett. 
{\bf 56}, 444 (1986).  

\bibitem{Ikoe93}
ORANGE collaboration, I. Koenig {\em et al.}, Z. Phys.  
{\bf A 346}, 153 (1993); see also 
W.~Koenig {\em et al.}, Phys. Lett.  {\bf B 218}, 12 (1989).

\bibitem{Tse91}
H. Tsertos, P. Kienle, S.M. Judge, K.~Schreckenbach,
Phys.  Lett.  {\bf B 266}, 259 (1991).

\bibitem{Lein96a}
U. Leinberger, Ph.D thesis, Tech. Univ. of Munich (1996), unpublished.

\bibitem{Shei96}
S. Heinz, Ph.D thesis (in preparation), Tech. Univ. of Munich.

\bibitem{APEX95}
APEX collaboration, I. Ahmad {\em et al.},
Phys. Rev. Lett.  {\bf 75}, 2658 (1995).

\bibitem{Gan96}
EPOS II  collaboration, R. Ganz {\em et al.},
Phys.  Lett.  {\bf B 389}, 4 (1996). 

\bibitem{Lein96b}
 ORANGE collaboration, U. Leinberger {\em et al.}, preprint 
 {\em (nucl-ex/9610001)}, accepted for publication in Phys. Lett. B.

\bibitem{Led78}
C. M. Lederer, V. S. Shirley, in: {\em Table of Isotopes},
John Wiley \& Sons Inc., NY, (1978). 

\bibitem{Bau96}
Jens Baumann, Ph.D thesis, University of Heidelberg (1996), unpublished;
and GSI report {\bf GSI--96--05}; and Jens Baumann, priv. comm.

\bibitem{Schl79}
P. Schl\"uter, G. Soff, 
Atomic and Nuclear Data Tables {\bf 24}, 509 (1979).

\bibitem{Hof}
Ch. Hofmann, G. Soff, J. Reinhardt, W. Greiner,
Phys. Rev. {\bf C 53}, 2313 (1996); and
Christian Hofmann, priv. comm.

\bibitem{Die96}
E. Ditzel et al., preprint (1996), to be published.

\bibitem{Soff81}
G. Soff, Z. Phys. {\bf A 303}, 189 (1981).

\end{thebibliography}
\end{document}